\documentclass{npw}

\usepackage{graphicx}
\usepackage{color}
\usepackage{mathrsfs}
\usepackage{dcolumn} %Align table columns on decimal point
\usepackage{bm}% bold math
\usepackage{epsfig}
\usepackage{wasysym}

\newcommand{\bsigma}{\mbox{\boldmath$\sigma$}}
\newcommand{\btau}{\mbox{\boldmath$\tau$}}

\begin{document}

\title{New Skyrme energy density functional for a better description of the Gamow-Teller Resonance \hspace{3mm}}

\author{X. Roca-Maza\email{xavier.roca.maza@mi.infn.it}\\
       \it INFN, sezione di Milano, via Celoria 16, I-20133 Milano, Italy\\
       G. Col\`o\\
       \it INFN, sezione di Milano, via Celoria 16, I-20133 Milano, Italy\\
       \it Dipartimento di Fisica, Universit\`a degli Studi di Milano, via Celoria 16, I-20133 Milano, Italy\\
        H. Sagawa\\
        \it Center for Mathematics and Physics, University of Aizu, Aizu-Wakamatsu, Fukushima 965-8560, Japan\\
        \it Nishina Center, Wako, Saitama 351-0198, Japan\\}
\pacs{21.60.Jz, 24.30.Cz}
\date{}
\maketitle

\begin{abstract}
We present a new Skyrme energy density functional (EDF) named SAMi \cite{roca12}. This interaction has been accurately calibrated to reproduce properties of doubly-magic nuclei and infinite nuclear matter. The novelties introduced in the model and fitting protocol of SAMi are crucial for a better description of the Gamow-Teller Resonance (GTR). Those are, on one side, the two-component spin-orbit potential needed for describing different proton high-angular momentum spin-orbit splitings and, on the other side, the careful description of the empirical hierarchy and positive values found in previous analysis of the spin ($G_0$) and spin-isospin ($G_0^\prime$) Landau-Migdal parameters: $0 < G_0 < G_0^\prime$, a feature that many of available Skyrme forces fail to reproduce. When employed within the self-consistent Hartree-Fock plus Random Phase Approximation, SAMi produces results on ground and excited state nuclear properties that are in good agreement with experimental findings. This is true not only for the GTR, but also for the Spin Dipole Resonance (SDR) and the Isobaric Analog Resonance (IAR) as well as for the non charge-exchange Isoscalar Giant Monopole (ISGMR) and Isovector Giant Dipole (IVGDR) and Quadrupole Resonances (IVGQR). 
\end{abstract}

\section{Introduction}
\label{introduction}
The Skyrme Hartree-Fock (HF) plus Random Phase Approximation (RPA) approach is one of the successful techniques for the study of the ground state and excited state properties of nuclei \cite{ston07, paar07}. The Skyrme ansatz is also used in more elaborated theoretical frameworks that include higher-order nuclear correlations \cite{ref_GCM, ref_PVC}.  

The Skyrme HF+RPA approach enables an effective description of the nuclear many-body problem in terms of a local energy density functional. Specific drawbacks and problems exist \cite{erl11} in this type of functionals. Therefore, we need to understand the origin of such deficiencies and eventually solve them. One of these problems is the accurate determination of the spin-isospin properties of that functional. If achieved, it should lead to accurate predictions of the properties of Gamow Teller Resonance (GTR) \cite{ost92}, namely, the main focus of this work for its relevance in electron-capture in the core-collapse of supernov\ae \cite{beth90,lang08}; neutrino-induced nucleosynthesis \cite{byel07}; the study of double-$\beta$ decay \cite{avig08} and, if ever observed, for determining the neutrino mass in neutrinoless double-$\beta$ decay. GT matrix elements are also very useful for the calibration of detectors aiming to measure electron-neutrinos \cite{lass02}.  

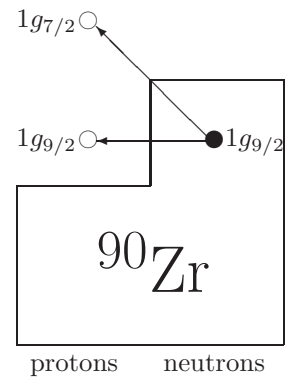
\begin{figure}[h!]
\vspace{1cm}
\begin{center}
\begin{picture}(100,100)(0,0)
\put(  5,-10){protons}
\put( 55,-10){neutrons}
\put(  0,  0){\line(1,0){100}}
\put(  0,  0){\line(0,1){ 60}}
\put(  0, 60){\line(1,0){ 50}}
\put( 50, 60){\line(0,1){ 40}}
\put( 50,100){\line(1,0){ 50}}
\put(100,  0){\line(0,1){100}}
\put( 75, 75){\vector(-1,1){45}}
\put( 75, 77){\vector(-1,0){45}}
\put( 70, 75){\CIRCLE}
\put( 23,120){\Circle}
\put( 23, 75){\Circle}
\put( 30, 20){\Huge ${}^{90}$Zr}
\put( 78, 75){$1g_{9/2}$}
\put(  0,120){$1g_{7/2}$}
\put(  0, 75){$1g_{9/2}$}
\end{picture}
\end{center}
\caption{Schematic picture of the most important single-particle transitions ($\sigma\tau_{-}$) involved in the Gamow Teller Resonance of ${}^{90}$Zr.}
\label{fig1}
\end{figure}

The sum of all possible transitions from the ground state $\vert 0\rangle$ to any possible excited state $\vert \nu\rangle$ give the linear response or GT strength function, 
\begin{equation}
R_{{\rm GT}^\pm} (E) = \sum_\nu \vert\langle \nu\vert\vert \hat{\mathcal{O}}_{\rm GT}^\pm\vert\vert 0\rangle\vert^2\delta(E-E_\nu),  
\end{equation}
where $\hat{\mathcal{O}}_{\rm GT}^\pm$ is the GT operator
\begin{equation}
\hat{\mathcal{O}}_{\rm GT}^\pm = \sum_{i=1}^A \bsigma(i)\tau_\pm(i),
\end{equation}
and $A$ is the mass number and $\bsigma$ and $\btau$ are the spin and isospin Pauli matrices, respectively. The dominant transitions will be those between spin-orbit partner levels as illustrated in the schematic picture in Fig.~\ref{fig1} for the case of ${}^{90}$Zr. If this is true also for other nuclei, it is easy to realize that spin-orbit splittings between nucleon levels above -- but still close to -- the Fermi surface will play a relevant role. In this respect, it is well known in the field that most of the Skyrme interactions overestimate the experimental spin-orbit splittings in heavy nuclei \cite{bend99}. 

GTR measurements shows that the Ikeda Sum Rule (ISR) given by $\int [R_{{\rm GT}^-}(E) - R_{{\rm GT}^+}(E)]{\rm d}E=3(N-Z)$ exhausts only a 60-70\% in the resonance region. To explain this quenching, it has been proposed that 2-particle 2-hole (2p-2h) correlations, or the coupling with the  $\Delta -$hole   excitation, have to be taken into account. Recent experimental analysis in $^{90}$Zr \cite{waka97} seems to indicate that most of the quenching (around 20\% of ISR) has to be attributed to 2p-2h coupling and that the role played by the $\Delta$ isobar is much less important. 

\section{Skyrme Hamiltonian and Random Phase Approximation}
\label{formalism}

The Skyrme interaction is a zero-range, velocity-dependent interaction that describe nucleons with space, spin and isospin variables. The Hamiltonian density is written as in Ref.~\cite{chab98},  
\begin{equation}
\mathcal{H} = \mathcal{K} + \mathcal{H}_0 + \mathcal{H}_3 + \mathcal{H}_{\rm eff} + \mathcal{H}_{\rm fin} + \mathcal{H}_{\rm SO} + \mathcal{H}_{\rm sg}+\mathcal{H}_{\rm Coul}
\label{hamiltonian}
\end{equation}
where $\mathcal{K}=\hbar^2\tau / 2m$ is the kinetic energy term where $\tau\equiv \sum_i^A {\bf p}_i^2$, $\mathcal{H}_0$ the zero range term, $\mathcal{H}_3$ the density dependent term, $\mathcal{H}_{\rm eff}$ the effective mass term, $\mathcal{H}_{\rm fin}$ the finite range term, $\mathcal{H}_{\rm SO}$ the spin-orbit term, $\mathcal{H}_{\rm sg}$ is a term due to the tensor coupling with spin and gradient and $\mathcal{H}_{\rm Coul}$ is the Coulomb term. The exchange part of the Coulomb term is calculated within the Slater approximation \cite{vaut72}. Specifically,  
\begin{eqnarray*}
\mathcal{H}_0       &=&  \frac{1}{4}t_0 \left[(2+x_0)\rho^2-(2x_0+1)(\rho_n^2+\rho_p^2)\right]\\
\mathcal{H}_3       &=&  \frac{1}{24}t_3\rho^\alpha\left[(2+x_3)\rho^2-(2x_3+1)(\rho_n^2+\rho_p^2)\right]\\
\mathcal{H}_{\rm eff} &=&  \frac{1}{8}\left[t_1(2+x_1)+t_2(2+x_2)\right]\tau\rho + \\
                    & & \frac{1}{8}\left[t_2(2x_2+1)-t_1(2x_1+1)\right](\tau_n\rho_n +\tau_p\rho_p)\\
\mathcal{H}_{\rm fin} &=& \frac{1}{32}\left\{\left[3t_1(2+x_1)-t_2(2+x_2)\right]({\bf\nabla}\rho)^2 - \right.\\
                    & & \left.\left[3t_1(2x_1+1)+ t_2(2x_2+1)\right]\left[({\bf\nabla}\rho_n)^2+({\bf\nabla}\rho_p)^2\right]\right\}\\
\mathcal{H}_{\rm SO}  &=& \frac{1}{2}W_0{\bf J\cdot\nabla}\rho + \frac{1}{2}W_0^\prime({\bf J\cdot_n\nabla}\rho_n+{\bf J_p\cdot\nabla}\rho_p\\
\mathcal{H}_{\rm sg}  &=& -\frac{1}{16}(t_1x_1+t_2x_2){\bf J}^2+\frac{1}{16}(t_1-t_2)({\bf J_n}^2+{\bf J_p}^2)
\label{hamiltionian-terms}
\end{eqnarray*}

The main differences with other Skyrme models available in the literature are that we include a two parameter, $W_0$ and $W_0^\prime$, spin-orbit potential \cite{rein95}. For $W_0=W_0^\prime$ one recovers the most standard form of this type of functional, for $W_0^\prime=0$ the spin-orbit potential mimicks relativistic calculations and for non-vanishing $W_0\neq W_0^\prime$, a mixed behavior is found. And that we also include $\mathcal{H}_{\rm sg}$ known as {\it central tensor} term or $J^2$ term. 

Pairing correlations, important for the description of open shell nuclei, and deformation are not included in our calculations since we study closed shell, double-magic spherical nuclei. The center-of-mass correction adopted in our calculations uses the factor  $m A / (A-1) $ instead of the bare nucleon mass $m$ in the kinetic energy term $\mathcal{K}$. This prescription accounts for a large part of the center-of-mass correction and it is implemeted in most of Skyrme-HF calculations. 

The discrete RPA method we adopt in our calculations is well-known in textbooks \cite{ring80,rowe80}. In our self-consistent approach, we build the residual interaction for the proton-proton, neutron-neutron and proton-neutron channels from the Skyrme-HF energy density functional. Then we solve fully self-consistently the RPA equations by means of the matrix formulation. For further details see Ref.~\cite{CPC} where, very recently, the code we have used for the calculations presented here has been published. 

\section{Fitting procedure}
\label{fit}

We present a non-relativistic functional of the Skyrme type \cite{roca12}, named SAMi for Skyrme-Aizu-Milano. It is as accurate as previous Skyrme models in the description of nuclear matter properties and of masses and charge radii of double-magic spherical nuclei. In addition, due to our fitting protocol, it improves the description of the GTR in medium and heavy mass nuclei with respect to previous models of the same type. 

For the minimization, we have performed a $\chi^2$ test (see Ref.~\cite{roca12} for further details) by means of a variable metric method included in the {\sc MINUIT} package of Ref.~\cite{jame96}. We have chosen the following set of data and $pseudo$-data for our fit: 
\begin{itemize}
\item[(i)] the binding energies of ${}^{40,48}$Ca, ${}^{90}$Zr, ${}^{132}$Sn and ${}^{208}$Pb and the charge radii of ${}^{40,48}$Ca, ${}^{90}$Zr and ${}^{208}$Pb;
\item[(ii)] the spin-orbit splittings of the 1$g$ and 2$f$ proton levels in ${}^{90}$Zr and ${}^{208}$Pb;
\item[(iii)] the fixed values for the Landau-Migdal parameters $G_0=0.15$ and $G_0^\prime=0.35$;
\item[(iv)] {\it pseudo}-data corresponding to the variational calculations of the energy per particle of uniform neutron matter at baryon density $\rho$ between $0.07$ fm${}^{-3}$ and $0.4$ fm${}^{-3}$ of Ref.~\cite{wiri88}.  
\end{itemize}
The novelties of our protocol that guide the fit to a better description of the GTR are twofold and due to points (ii) and (iii). Actually, the impact on the excitation energy and strength of the GTR of (ii) and (iii) have been studied in previous literature \cite{giai81,bend02,gang10}. The importance of (ii) has been already discussed in the introduction. The hierarchy and values that we have taken for the spin and spin-isospin Landau-Migdal parameters in point (iii) have been empirically suggested by the study of Ref.~\cite{waka07} (find more details also in Ref.~\cite{roca12}). 

On the other side, point (i) has allowed us to determine the saturation energy ($e_\infty$), density ($\rho_\infty$) and incompressibility ($K_\infty$) of symmetric nuclear matter -- constrained to be $240\pm20$ MeV by an analysis of a large set of Skyrme interactions \cite{colo04} -- and point (iv) has been helpful in driving the magnitude ($J$) and slope ($L$) of the nuclear symmetry energy at nuclear saturation density towards reasonable values \cite{trip08,cent09,tsan12,piek12,prex12,roca11}. In Tables 1 and 2 of Ref.~\cite{roca12} the references for the used data and {\it pseudo}-data with the corresponding adopted errors, partial contributions to the $\chi^2$, and the number of data points used in the fit can be found (Table 1); and the SAMi parameter set and some saturation properties with the estimated standard deviations (Table 2). 

\section{Results}
\label{results}

In this section we will show the results of SAMi as compared to other interactions and experimental data for the GTR in  ${}^{48}$Ca, ${}^{90}$Zr and ${}^{208}$Pb and the Spin Dipole Resonance (SDR) in ${}^{90}$Zr and ${}^{208}$Pb. Other important results such as binding energies, charge radii and the comparison of pure neutron and symmetric matter equations of states (EoSs) predicted by SAMi with state-of-the-art Brueckner-Hartree-Fock and the fitted variational calculations of Ref.~\cite{wiri88} can be found in Ref.~\cite{roca12}. In addition to that, we have also tested that SAMi EoS is stable against spin and spin-isospin instabilities \cite{gang10} up to a baryon density of more than four times saturation density, well above the region important for the description of finite nuclei and enough for the study of uniform neutron-rich matter in neutron stars. 

The predictions of SAMi for important non charge-exchange excitations such as the Isoscalar Giant Monopole Resonance (ISGMR), Isovector Giant Dipole Resonance (IVGDR) and the Isovector Giant Quadrupole resonance (IVGQR) in ${}^{208}$Pb are accurate. Specifically, the excitation ($E_x$) -- or centroid ($E_{\rm c}$)--  energies as predicted by our HF+RPA calculations (experimental values) for these resonances are $E_{\rm c}({\rm GMR})= 14.48$ MeV ($E_{\rm c}({\rm GMR})=14.24\pm0.11$ MeV \cite{youn99}), $E_{\rm c}({\rm GDR})= 13.95$ MeV ($E_{\rm c}({\rm GDR})=13.25\pm 0.10$ MeV \cite{ryez02}) and $E_x({\rm GQR})= 23.1$ MeV ($E_x({\rm GQR})=23.0\pm 0.2$ MeV \cite{hens11}), respectively. The corresponding energy weighted sum rules agree well with available experimental data.    

\begin{figure}[t!]
\includegraphics[width=0.9\linewidth,clip=true]{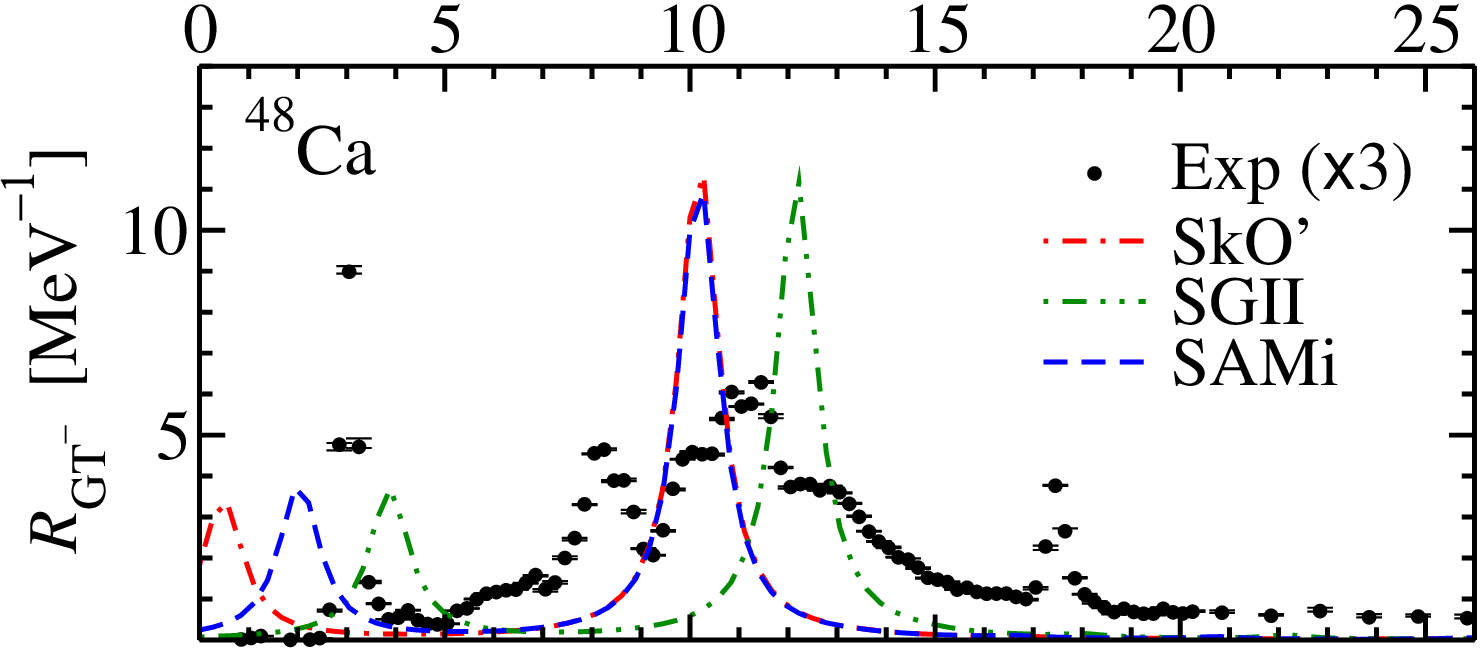}\\ 
\includegraphics[width=0.9\linewidth,clip=true]{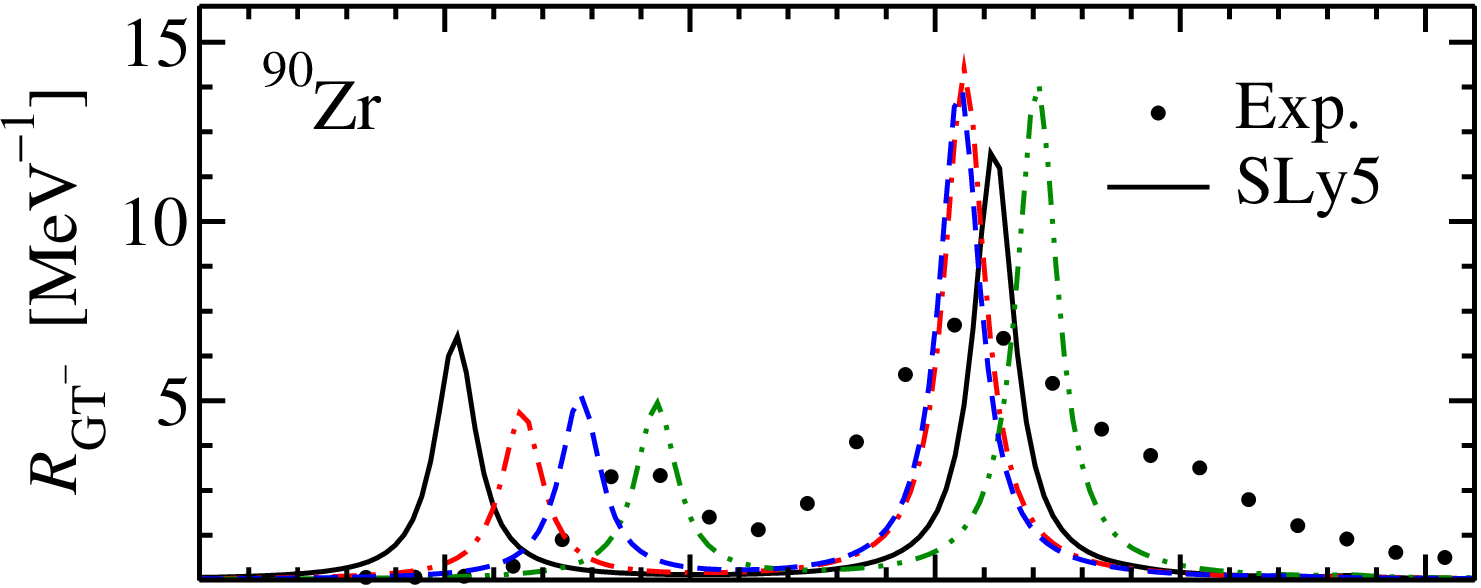}\\ 
\includegraphics[width=0.9\linewidth,clip=true]{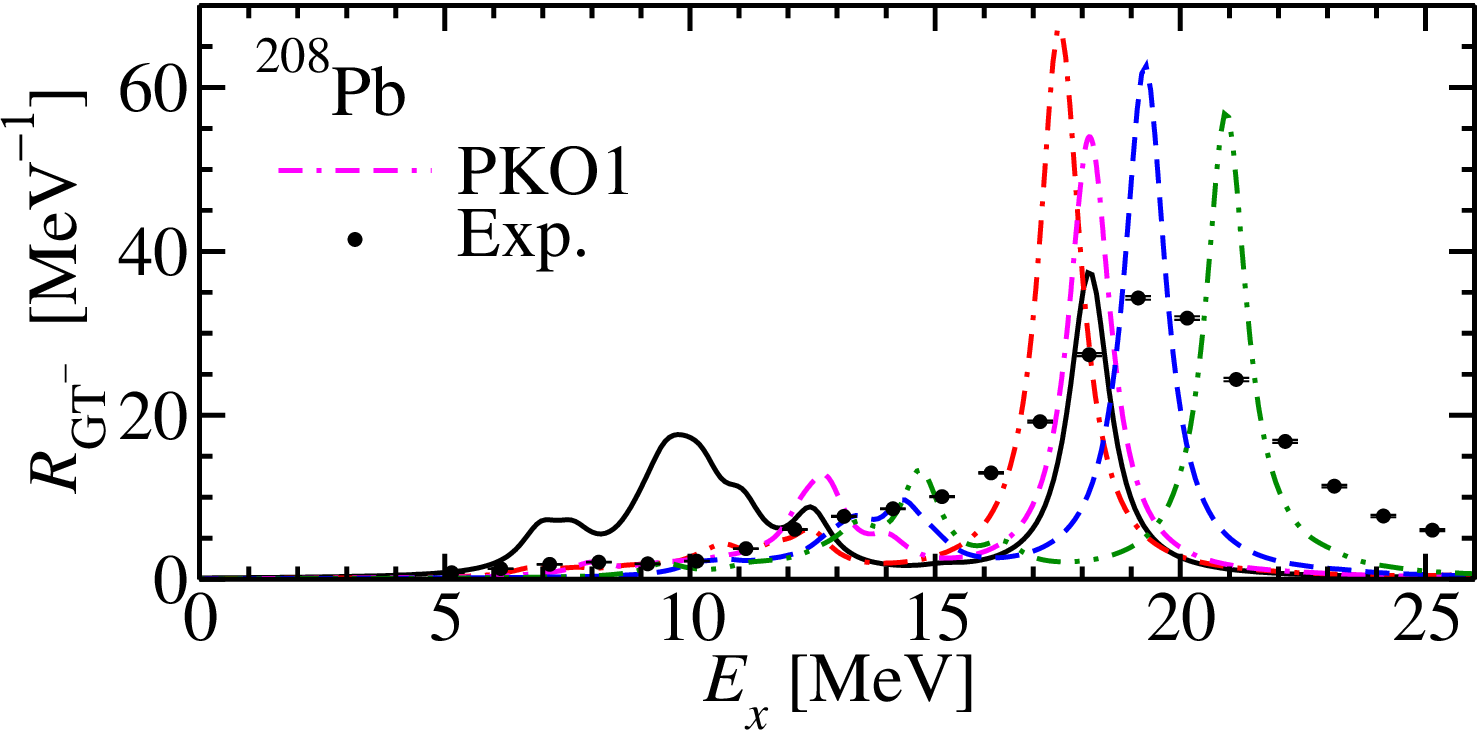}
\caption{GT strength distributions in ${}^{48}$Ca (upper panel), ${}^{90}$Zr (middle panel) and ${}^{208}$Pb (lower panel) as measured in the experiment \cite{yako09,waka97,kras01,akim95,waka12} and predicted by SLy5 \cite{chab98}, SkO' \cite{rein99}, SGII \cite{giai81} and SAMi forces. For the case of ${}^{208}$Pb we also show the predictions of PKO1 \cite{lian08}.}
\label{gt}  
\end{figure}

The earliest attempt to give a quantitative description of the GTR was provided by the Skyrme SGII interaction \cite{giai81}. Later on, an accurate functional for the predictions of finite nuclei and charge-exchange resonances was proposed: namely SkO' \cite{rein99}. Relativistic mean-field and relativistic HF calculations of the GTR have also become available meanwhile \cite{paar04,lian08}. For this reason, we show in Fig.~\ref{gt}, the results for the Gamow-Teller strength distributions in ${}^{48}$Ca (upper panel), ${}^{90}$Zr (middle panel) and ${}^{208}$Pb (lower panel) as measured in the experiment \cite{yako09,waka97,kras01,akim95,waka12} and as predicted the non-relativistic SkO' \cite{rein99}, SGII \cite{giai81} and SAMi interactions obtained within our HF+RPA calculations. We also show the predictions of SLy5 \cite{chab98} since it is of common use in the field and because our fitting protocol is similar to the one used to obtain this interaction and the predictions of PKO1 \cite{lian08} for the case of ${}^{208}$Pb since it is based on a relativistic framework.    

In the upper panel of Fig.~\ref{gt}, we show the experimental data as well as the predictions of the SAMi, SGII and SkO' functionals for ${}^{48}$Ca (the SLy5 result is not shown because RPA calculations produce instabilities, i.e., imaginary energies). The excitation energy found in the experiment for the high-energy peak, $E_x^{\rm exp}=10.5$ MeV, and that predicted by SAMi, $E_x^{\rm SAMi}=10.2$ MeV, are in very good agreement. For the low-energy peak, the agreement of the excitation energy found in the experiment, $E_x^{\rm exp}=3.0$ MeV, with that of SAMi, $E_x^{\rm SAMi}=2.0$ MeV, is also good. The corresponding \% of the ISR exhausted by the main peak between 5 and 17 MeV, is around 46\% in the experiment and 71\% in our theoretical calculations with SAMi (note that RPA does not include 2p-2h couplings). The overall description of the GTR in ${}^{48}$Ca by SAMi improves the results obtained with SGII and SkO'.   

The prediction of the SAMi interaction in the case of ${}^{90}$Zr (middle panel of Fig.~\ref{gt}) is even better than in the case of $^{48}$Ca. The excitation energy and \% of the ISR exhausted by the high- and low-energy peaks in the experimental data (SAMi) are, respectively, $E_x^{\rm exp}=15.8\pm0.5$ MeV and 57\% ($E_x^{\rm SAMi}=15.5$ MeV and 70\%) between 12 and 30 MeV and $E_x^{\rm exp}=9.0\pm0.5$ MeV and 12\% ($E_x^{\rm SAMi}=7.8$ MeV and 27\%)  between 3 and 12 MeV. Again, the best overall picture is given by SAMi when compared with the other Skyrme interactions SGII, SkO' and SLy5.

With unprecedented accuracy in HF+RPA calculations, the SAMi functional perfectly reproduces the excitation energy of the experimental GTR in ${}^{208}$Pb \cite{akim95} (lower panel of Fig.~\ref{gt}): $E_x^{\rm exp}=19.2\pm0.2$ MeV and $E_x^{\rm SAMi}=19.3$ MeV. We also compare our results with the predictions of SGII, SLy5, SkO' and PKO1. None of them reproduces the excitation energy of the GTR in this nucleus within the experimental accuracy. 
%For the sake of clarity, it is important to notice that, opposite to SLy5 and SGII, the spin-orbit parameters ($W_0$ and $W_0^\prime$) are not fixed to be equal in the SAMi and SkO' interactions, that $G_0'$ is too large for the case of SGII \cite{giai81} and that SAMi gives a reasonable description of the proton high-angular spin-orbit splittings close to the Fermi level. 

%
\begin{figure}[t]
\includegraphics[width=0.9\linewidth,clip=true]{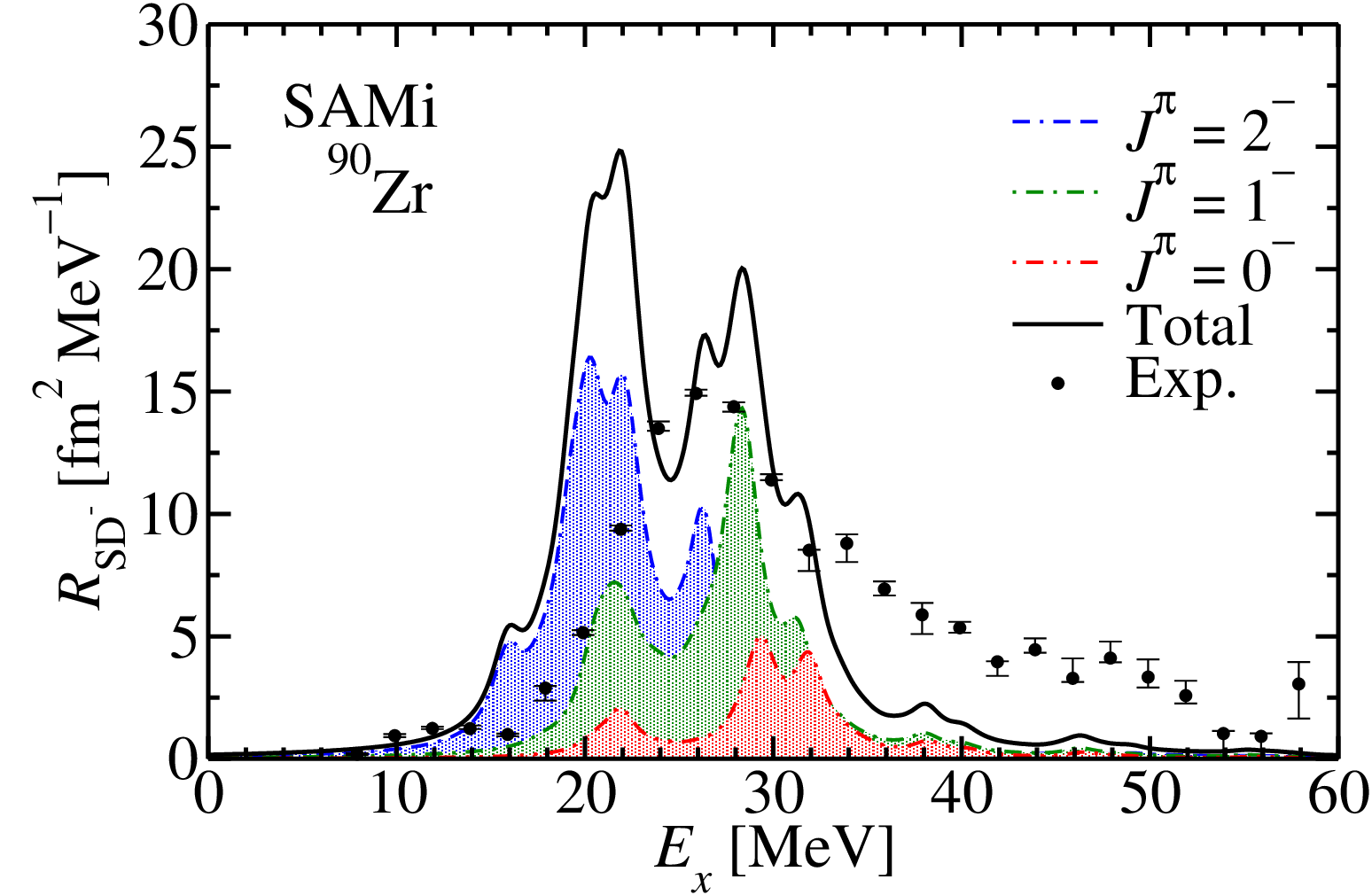}\\ 
\caption{Spin Dipole strength distribution for ${}^{90}$Zr in the $\tau_{-}$ channel as a function of the excitation energy $E_x$ measured in the experiment \cite{yako06} and predicted by SAMi. Multipole decomposition is also shown. A Lorentzian smearing parameter equal to 2 MeV is used.}
\label{sdr-zr}
\vspace{-0.5cm}
\end{figure}

The SAMi reliability for the calculation of other charge-exchange resonances such as the SDR or the Isobaric Analog Resonance in some doubly-magic spherical nuclei have been tested in Ref.~\cite{roca12}. Here, we show the results for the SDR in ${}^{90}$Zr and ${}^{208}$Pb. The operator used for the RPA calculations is,
\begin{equation}
\sum_{i=1}^A \sum_{M}\tau_\pm(i)r_i^{L}\left[Y_{L}(\hat{r}_i)\otimes\bsigma(i)\right]_{JM}
\end{equation} 
and, as it is shown in both figures, it connects single particle states differing by a total angular momentum: $J^{\pi} = 0^-, 1^-$ and $2^-$. The sum rule, 
\begin{equation}
\int [R_{{\rm SD}^-}(E) - R_{{\rm SD}^+}(E)]{\rm d}E=\frac{9}{4\pi}(N\langle r_n^2\rangle-Z\langle r_p^2\rangle),
\end{equation}
 is completely exhausted in our calculations, 99.99\% in the case of ${}^{90}$Zr and 100\% in the case of ${}^{208}$Pb. 

\begin{figure}[t]
\includegraphics[width=0.9\linewidth,clip=true]{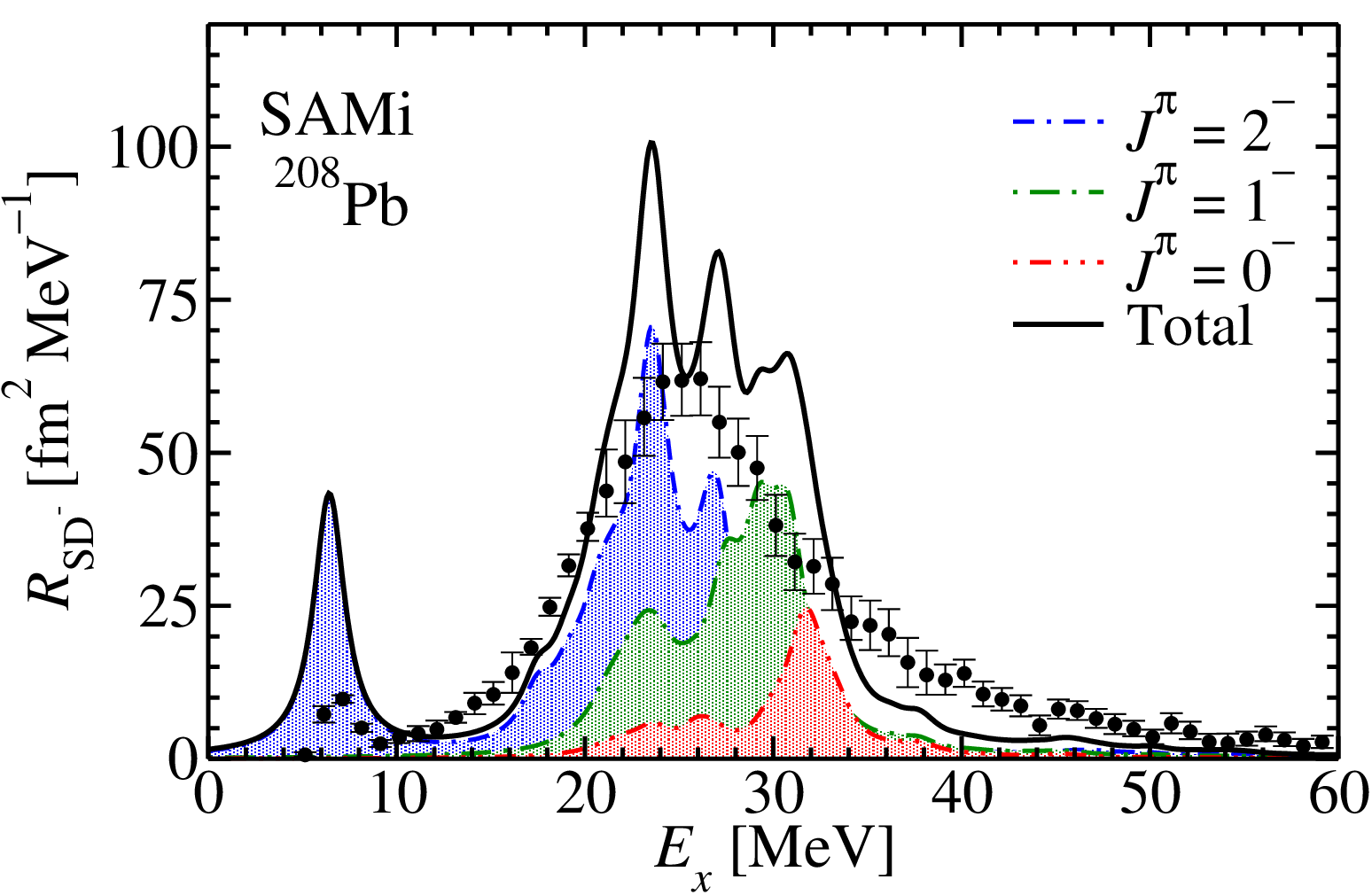}\\ 
\caption{SDR strength distributions for ${}^{208}$Pb in the $\tau_{-}$ channel from experiment \cite{waka12} and SAMi calculations. Total and multipole decomposition of the SDR strength are shown. A Lorentzian smearing parameter equal to 2 MeV is used.}
\label{sdr-pb}
\vspace{-0.5cm}
\end{figure}

In Figs.~\ref{sdr-zr} and \ref{sdr-pb} the SDR in ${}^{90}$Zr and ${}^{208}$Pb are compared with the experimental results for the total and multipole decomposition ($J^{\pi} = 0^-, 1^-$ and $2^-$) contributions. The overall agreement is noticeable. For the case of ${}^{90}$Zr, the experimental value for the sum rule is $148\pm12$ fm${}^{2}$ \cite{yako06} and the SAMi prediction is $150$ fm${}^{2}$. The total and multipole decomposition ($J^{\pi} = 0^-, 1^-$ and $2^-$) of the experimental data \cite{waka12} for the integral of $R_{{\rm SD}^-}$ in the case of ${}^{208}$Pb are 1004$^{+24}_{-23}$ fm${}^{2}$, 107$^{+8}_{-7}$ fm${}^{2}$, 450$^{+16}_{-15}$ fm${}^{2}$ and 447$^{+16}_{-15}$ fm${}^{2}$; and for the SAMi predictions 1224 fm${}^{2}$, 158 fm${}^{2}$,  423 fm${}^{2}$ and 643 fm${}^{2}$, respectively. 

\section{Summary and conclusions}
\label{conclusions}
The new Skyrme interaction presented here and in more detail in Ref.~\cite{roca12} accounts for the most relevant quantities in order to improve the description of charge-exchange nuclear resonances, i.e., the hierarchy and value of the spin and spin-isospin Landau-Migdal parameters and the proton spin-orbit splittings of different high angular momentum single-particle levels close to the Fermi surface. As a proof, the GTR in $^{48}$Ca, $^{90}$Zr and $^{208}$Pb and other charge-exchange resonances \cite{roca12} are predicted with high accuracy by SAMi without compromising the description of other nuclear observables. 

\begin{ack}
We are very grateful to T. Wakasa, H. Sakai and K. Yako for providing us with the experimental data on the presented charge-exchange resonances for ${}^{48}$Ca, ${}^{90}$Zr and ${}^{208}$Pb and to H. Liang for providing us with the predictions of the PKO1 interaction for ${}^{208}$Pb. The support of the Italian Research Project ``Many-body theory of nuclear systems and implications on the physics of neutron stars'' (PRIN 2008) is acknowledged.
\end{ack}

%\section*{References}

\end{document}